\documentclass{article}
\usepackage{arxiv}
\usepackage{graphicx,amsmath,amsfonts,amscd,amsthm,amssymb,pstricks}
\usepackage{color}
\usepackage{dcolumn}
\usepackage{bm}
\usepackage{longtable}
\usepackage{mathrsfs}
\usepackage{textcomp}
\usepackage{esvect}
\usepackage{float}
\usepackage{mathtools,amsfonts}
\usepackage[USenglish,british]{babel}
\usepackage{environ}
\usepackage{xifthen}
\usepackage{xargs}		
\usepackage{hyperref}
\usepackage{physics,mathtools}
\usepackage[separate-uncertainty = true]{siunitx}
\usepackage{multirow}
\usepackage{comment}
\usepackage{enumerate}
\usepackage{appendix}
\usepackage{tikz}	
\usetikzlibrary{arrows, calc, matrix, patterns, decorations.markings, shapes, decorations.pathmorphing, shadows.blur}
\usepackage[utf8]{inputenc}
\usepackage{enumitem}

\newcommand{\ie}{i.e.\ }

\newcommand{\eg}{e.g.\ }

\newcommand{\ham}{\mathcal{H}}
\newcommand{\map}{\mathcal{M}}
\newcommand{\eps}{\varepsilon}

\newcommand{\epsm}{\eps_\text{m}}

\newcommand{\epsmf}{\eps_\text{m,f}}
\newcommand{\epsh}{\eps_\text{h}}

\title{Analysis of adiabatic trapping phenomena for quasi-integrable area-preserving maps in the presence of time-dependent exciters}

\author{A. Bazzani\\
Dipartimento di Fisica e Astronomia, Universit\`a di Bologna and INFN Bologna, via Irnerio 46, Bologna, Italy\\
\And
F. Capoani\\
Dipartimento di Fisica e Astronomia, Universit\`a di Bologna and INFN Bologna, via Irnerio 46, Bologna, Italy\\
Beams Department, CERN, Esplanade\ des Particules 1, 1211 Geneva 23, Switzerland\\
\And
M. Giovannozzi\thanks{Corresponding author: massimo.giovannozzi@cern.ch}\\
Beams Department, CERN, Esplanade\ des Particules 1, 1211 Geneva 23, Switzerland}

\begin{document}
\maketitle

\begin{abstract}
In this paper, new results concerning the phenomenon of adiabatic trapping into resonance for a class of quasi-integrable maps with a time-dependent exciter are presented and discussed in detail. The applicability of the results about trapping efficiency for Hamiltonian systems to the maps considered is proven by using perturbation theory. This allows determining explicit scaling laws for the trapping properties. These findings represent a generalization of previous results obtained for the case of quasi-integrable maps with parametric modulation as well as an extension of the work by Neishtadt \textit{et al.} on a restricted class of quasi-integrable systems with time-dependent exciters. 
\end{abstract}



\section{Introduction}

The adiabatic theory for Hamiltonian systems is a key breakthrough towards an understanding of the effects of slow parametric modulation on the dynamics. The concept of adiabatic invariant allows the long-term evolution of the system to be predicted and the fundamental properties of the action variables to be highlighted upon averaging over the fast variables~\cite{arnold,chirikov}. The theory has been well developed for systems with one degree of freedom~\cite{neish1976,neish1985,an6,dv_im,an10}, but the extension of some analytical results to multi-dimensional systems or to symplectic maps~\cite{ab_fb_gt_AIP} has to cope with the issues generated by small denominators and the ubiquitous presence of resonances in phase space~\cite{via1,via3}. For these reasons, such an extension is still an open problem. 

Recently, the possibility of a controlled manipulation of the phase space by means of an adiabatic change of a parameter opened the road-map to new applications in accelerator and plasma physics~\cite{an4,sridhart,bvc4,escande2016,mte1,mte2}. In particular, the adiabatic transport performed by means of nonlinear resonance trapping allows to manipulate a charged particle distribution so to minimize the particle losses during the beam extraction process in a circular accelerator. Furthermore, the control of the beam emittance can be obtained by a similar approach~\cite{mte2017,mte3,mte4}. The experimental procedures~\cite{mte2017,mte3,mte4} require a very precise control of the efficiency of the adiabatic trapping into resonances~\cite{an10,an9,add2}, as well as of the phase-space change during the adiabatic transport when the parametric modulation is introduced by means of an external perturbation. All these processes can be represented by multi-dimensional Hamiltonian systems or symplectic maps~\cite{bazzanietal}. 

In this paper, we consider the problem of obtaining an accurate estimate of the resonance-trapping efficiency and of the phase-space transport for a given distribution of initial conditions in the case of polynomial symplectic maps when a time-dependent periodic perturbation is present. The perturbation frequency and amplitude are adiabatically changed. We show that the concept of interpolating Hamiltonian can be applied to derive the scaling laws of the main parameters of the map, \ie the perturbation amplitude, and the nonlinearity coefficients. In this way, we obtain explicit analytical estimates for the trapping and transport efficiencies, thus generalizing the analytical results obtained for Hamiltonian systems. The accuracy of the proposed estimates has been verified by means of extensive numerical simulations of different study cases. Furthermore, we  study the limits of the adiabatic approximation for the observed phenomena, and of the validity of our results. Note that the modulation of the external perturbation parameters is realized according to procedures that could open the way to new applications in the field of accelerator physics, in view of devising novel beam manipulations. 

The paper is organized as follows: in Section~\ref{sec:theory} we recall some theoretical results of the adiabatic theory that are needed to measure the resonance-trapping phenomenon, and we introduce the map models. In Section~\ref{sec:trapping} we perform a detailed analysis of the phase-space evolution during the trapping process, whereas in Section~\ref{sec:simresults} we discuss the results of detailed numerical simulations about the evolution of a particle distribution, comparing the dynamics of the interpolating Hamiltonian with that of the corresponding symplectic maps. In section~\ref{sec:complex}, a more complex model is presented and discussed to show that in spite of its features, the theory works well in generic systems. Finally, some conclusions are drawn in Section~\ref{sec:conc}, and some detailed computations of the the perturbation-theory calculations for a Hamiltonian system with a time-dependent exciter and the minimum action for which trapping occurs are reported in  Appendix~\ref{sec:appA}, and~\ref{sec:app_rmin}, respectively. 
 
\section{Theory} \label{sec:theory}
\subsection{Generalities}
Phenomena occurring when a Hamiltonian system is slowly modulated have been widely studied in the framework of adiabatic theory~\cite{neish1975, neish1985}. As the modulation of the Hamiltonian changes the shape of the separatrices in phase space, the trajectories can cross separatrices and enter into different stable regions that are associated with nonlinear resonances. The probability of the separatrix crossing, which is described by a random process in the adiabatic limit, can be computed as well as the change of adiabatic invariant due to the crossing~\cite{neish1975, neish1985}.

\begin{figure}
    \begin{center}
    \includegraphics[trim=45truemm 185truemm 45truemm 45truemm,width=0.7\columnwidth,clip=]{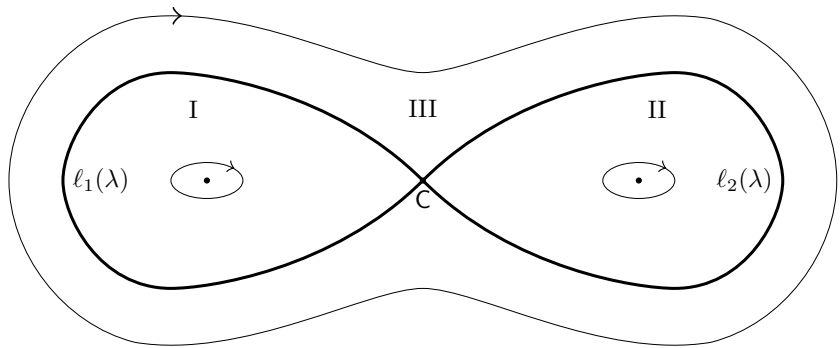}
    \end{center}
    \caption{A generic phase-space portrait divided into three regions ($\mathrm{I}$, $\mathrm{II}$, $\mathrm{III}$) by separatrices $\ell_1(\lambda)$ and $\ell_2(\lambda)$.}
    \label{fig:genericphsp}
\end{figure}

Let us consider a Hamiltonian $\ham(p,q,\lambda=\epsilon \, t) \, , \,\, \epsilon \ll 1$, where the parameter $\lambda$ is slowly modulated and whose phase space is sketched in Fig.~\ref{fig:genericphsp}. If we consider an initial condition that lies in Region~$\mathrm{III}$, the transition probability into Region~$\mathrm{I}$ or $\mathrm{II}$ of phase space is given by~\cite{neish1975}

\begin{equation} 
  \mathcal{P}_{\mathrm{III}\to \mathrm{I}} = \frac{\Theta_\mathrm{I}}{\Theta_\mathrm{I}+\Theta_\mathrm{II}}\,\qquad \mathcal{P}_{\mathrm{III}\to \mathrm{II}} = 1 - \mathcal{P}_{\mathrm{III}\to \mathrm{I}} \, ,
  \label{eq:neish}
\end{equation}
where
\begin{equation}
 \Theta_i = \dv{A_i}{\lambda}\eval_{\tilde\lambda} = \oint_{\partial A_i}\dd t\, \pdv{\ham}{\lambda}\eval_{\tilde\lambda}\qquad i=\mathrm{I},\,\mathrm{II} \, , 
 \label{eq:prob}
\end{equation}
with $A_i$ the area of the region $i$, $\partial A_i$ the boundary of region $i$, and $\tilde\lambda$ the value of $\lambda$ when the separatrix is crossed. We remark that in case $\mathcal{P}_{\mathrm{III}\to i} < 0$, then $\mathcal{P}_{\mathrm{III}\to i}$ is set to zero, whereas when $\mathcal{P}_{\mathrm{III}\to i}>1$ then $\mathcal{P}_{\mathrm{III}\to i}$ is set to unity.

When a separatrix is crossed, the adiabatic invariant $J$ is expected to change according to the area difference between the two regions at the crossing time, so that just after the crossing into a region of area $A$, we have $J=A/(2\pi)$. However, this value is exact only if the modulation is perfectly adiabatic, \ie it is infinitely slow. A correction to the value of the new action can be found following~\cite{neish1985}, and it is a random value, whose distribution depends on the random variable $\xi_i = |h|/(\epsilon \, \Theta_i)$, $h$ being the difference between the energy of a particle and that of the separatrix. The value of $\xi$ can be considered a random variable in the adiabatic limit, due to its extreme sensitivity to the initial condition of the action. 

The adiabatic trapping into resonances has been studied in various works~\cite{neish1975, Neishtadt2013} to show the possibility of transport in phase space when some system's parameters are slowly modulated. This phenomenon suggests possible applications in different fields and, in particular, in accelerator physics where the Multi-Turn Extraction has been proposed~\cite{mteprl} and successfully made an operational beam manipulation at the CERN Proton Synchrotron~\cite{Borburgh:2137954,mte2017}. In this case, an extension of the results of adiabatic theory to quasi-integrable area-preserving maps has been considered, and analogous probabilities~\eqref{eq:neish} to be captured in a resonance can be computed~\cite{bazzanietal} when the Poincar\'e--Birkhoff theorem~\cite{arnoldergo} can be applied to prove the existence of stable islands in phase space. The properties of such resonance islands for polynomial H\'enon-like maps~\cite{henon} have been studied in~\cite{yellowreport} and the possibility of performing an adiabatic trapping into a resonance has been studied by modulating the linear frequency at the elliptic fixed point~\cite{bazzanietal}. In this paper we extend the previous results by considering the adiabatic trapping in area-preserving maps when we introduce a time-modulated external sinusoidal term in the dynamics, with amplitude proportional to $q^m \, , m \in \mathbb{N}\, , m 
\geq 1 $, and whose frequency is adiabatically changed to cross a resonance with the unperturbed frequency of the system. This external forcing is particularly relevant for applications, as it mimics the effect of a transverse kicker on the charged particle dynamics in a circular accelerator (see, \eg~\cite{PhysRevSTAB.5.054001,PhysRevSTAB.8.024001,PhysRevSTAB.8.024401,PhysRevSTAB.11.084002,PhysRevSTAB.16.071002,PhysRevAccelBeams.19.054001} and references therein). Furthermore, it would allow extending the possibility to perform an efficient beam trapping into stable islands even when the unperturbed frequencies of the system cannot be modulated. This might be the case \eg in a circular particle accelerator in the presence of space charge effects that impose a special choice of linear tunes. 

\subsection{The models used} 

We consider a H\'enon-like symplectic map of the form 

\begin{equation} 
    \begin{split}
	    \mathcal{M}_{\ell,m}:\,	& \begin{pmatrix} q_{n+1}\\p_{n+1} \end{pmatrix} = R(\omega_0) \times \\
	    \times & \begin{pmatrix}q_n \\ p_n - \sum_{j>2} k_j q_n^{j-1} - q^{\ell - 1} \epsm \cos \omega \, n\end{pmatrix} \, , 
	\end{split}
        \label{eq:map}
\end{equation}
where $R(\omega_0)$ is a rotation matrix of an angle $\omega_0$, $n$ is the iteration number, $\ell \in \mathbb{N}$, and the dynamics is perturbed by a modulated kick of amplitude $\epsm$ whose frequency $\omega$ is close to a resonance condition $\omega = m\, \omega_0 +\delta \, , \, \delta \ll 1$. We remark that when $\ell=1$, the fixed point at the origin of the unperturbed system becomes an elliptic periodic orbit of period $2\pi/\omega$, and the linearized frequencies depend on the perturbation strength, so that they are adiabatically modulated. This is not the case when $\ell \ge 2$,  which is also interesting for applications. We shall consider explicitly these two cases.

The Birkhoff Normal Form theory allows a relationship between the map of Eq.~\eqref{eq:map} and the Hamiltonian~\cite{yellowreport}

\begin{equation}
\ham_{\ell,m} (p,q,t) = \omega_0 \frac{q^2+p^2}{2} + \sum_{j>2} \hat k_j \frac{q^j}{j} + \epsh \frac{q^\ell}{\ell} \cos\omega \, t 
\label{eq:ham}
\end{equation} 
to be established.

In particular, from Eq.~\eqref{eq:map} we can derive the Normal Form Hamiltonian
\begin{equation}
	\begin{split}
	\hat \ham_{\ell,m} (\hat J,\hat \theta) = \omega_0 \hat J &+ \sum_{j>2}  \Omega_{j-1}(k_j) \hat J^\frac{j-1}{2} + \\
	&+ \epsm \, c_{\ell,m} \, \hat{J}^{m/2} \cos (m \,\hat \theta - \omega \, t) \, , 
	\end{split}
\label{eq:ham_nf}
\end{equation} 
where $\Omega_j(k_j)$ are detuning terms, obtained from the non-resonant Normal Form, while $c_{\ell,m}$ is the Fourier component of the $m$th harmonic, which is the only one remaining when $\omega\approx m\omega_0$. We remark that the dependence of $\Omega_{j-1}$ on the $k_j$ is in the form of a polynomial, and the scaling laws have been derived in~\cite{yellowreport}. For instance, in the case of $\Omega_2$ one has $\Omega_2=a\, k_3^2 + b\, k_4$, where
\begin{equation}
a = -\frac{1}{16}\qty[3\cot(\frac{\omega_0}{2})+\cot(\frac{3\, \omega_0}{2})] \, . 
\label{eq:detuning_sex}
\end{equation}

If the Hamiltonian~\eqref{eq:ham} is averaged on the resonance, one obtains an expression of the same form as Eq.~\eqref{eq:ham_nf}, \ie
\begin{equation}
	\begin{split}
	\ham_{\ell,m}(J,\theta) = \omega_0 J &+ \sum_{j>2} \, \frac{\omega_{j-1}(\hat k_j)}{2} J^\frac{j-1}{2} \\ &+ \epsh \, c_{\ell,m}\,J^{m/2}\cos(m\, \theta - \omega \, t) \, ,
\end{split}	
\label{eq:ham_av}
\end{equation}
where we have the same $c_{\ell,m}$ Fourier coefficient that appears from the expansion of $\cos^\ell\theta\cos\omega t$ on the resonant harmonic, and $\omega_J$ is a polynomial function of the $k_j$.

From these two expressions, it is possible to evaluate the relation between the corresponding parameters, \ie $k_j, \hat{k}_j$ and $\epsm, \epsh$, which enables applying the analytical results valid for the Hamiltonian system to the corresponding polynomial symplectic map of the form~\eqref{eq:map}. 

Let $(\hat{J}, \hat{\theta})$ be the action-angle variables for the map defined by the perturbation series. The frequency of the angle dynamics can be written in the form~\cite{yellowreport}
\begin{equation}
\Omega(\hat J)=\omega_0+\sum_{j>2} \Omega_{j-1}( k_j) \, \hat J^\frac{j-1}{2} \, , 
\end{equation}
while from the Hamiltonian~\eqref{eq:ham}, which is expressed in terms of the action-angle variables $(J,\theta)$, we obtain
\begin{equation}
\left \langle \pdv{\ham_{\ell,m}}{J} \right \rangle_\theta =\omega_0 + \sum_{j>2} \omega_{j-1} (\hat{k}_j)\, J^\frac{j-1}{2} \, ,
\end{equation}
where $\langle \,\, \rangle_\theta$ stands for the average over the variable $\theta$.

If a single detuning term of order $j$ is considered and a single term $k_l$ is present in the system under study, and $\Omega_{j-1}=\Omega_{j-1, 0} k_l^r$ (and similarly for $\omega_{j-1}$), then one can assume that $k_l=\hat{k}_l$ and only the action variables need to be re-scaled to ensure the same frequency variation with action for the two systems under consideration. If $\hat J = \kappa \, J$, then from the expression
\begin{equation}
\omega_0 + k_l^r\, \Omega_{j-1,0} \, \kappa^{\frac{j-1}{2}} \, J^{\frac{j-1}{2}} = \omega_0 + k_l^r\, \omega_{j-1,0} \, J^{\frac{j-1}{2}}
\end{equation}
we derive
\begin{equation}
\kappa = \left ( \frac{\omega_{j-1,0}}{\Omega_{j-1,0}} \right )^\frac{2}{j-1} \, . 
\end{equation}

In the more general case in which several detuning terms are considered, but a single term $k_l$ is present, we have $\Omega_{j-1}(k_l)=k_l^{r_{l,j}} \Omega_{j-1,0}$ (and similarly for $\omega_{j-1}(k_l)$). The approach consists of re-scaling $\hat{k}_l=\kappa_{l,j} k_l$ instead of the action, which would then give the solution
\begin{equation}
\kappa_{l,j} = \left ( \frac{\omega_{j-1,0}}{\Omega_{j-1,0}} \right )^{1/r_{l,j}} \, ,
\end{equation}
thus making the frequency variation the same for both systems also in this case.

For $\ell=1$ and $m=3$, the outlined approach gives 
\begin{equation}
\kappa = \frac{\omega_{2,0}}{\Omega_{2,0}} \, , 
\end{equation}
and we still need to match the strength of the time-dependent perturbation, which is proportional to $\epsm \, c_{1,3} \, \hat J^{3/2} = \epsm \, c_{1,3} \, \kappa^{3/2} \, J^{3/2} = \epsh \, c_{1,3} \, J^{3/2}$, so that
\begin{equation}
\frac{\epsh}{\epsm} = \qty(\frac{\omega_{2,0}}{\Omega_{2,0}})^{3/2} \, .
\label{eq:epsratio}
\end{equation}

In the numerical simulations, we set $k_3=1$ and $\omega_0/(2\pi)=0.1713$, finding $\epsh/\epsm \approx 3.92$ from the computation of the Fourier coefficient $c_{1,3}$ by means of the perturbation theory, which is found in Appendix~\ref{sec:appA}).

Performing analogous computations for $\ell=2$ and $m=3$ (see Appendix~\ref{sec:appA}), the coefficient $c_{2,3}$ can be computed and by comparing it with the corresponding coefficient $c_{1,3}$ for the case $\ell=1$ one determines the different scale of the perturbation strength in the two cases, namely

\begin{equation} 
\frac{c_{1,3}}{c_{2,3}} = \frac{19}{62} \frac{k^2_3}{\omega_0}\, .
\end{equation}

According to the parameters used in the simulations ($k_3=1$ and $\omega_0/ 2\pi = 0.1713$) we obtain $c_{1,3}/c_{2,3} = 0.284$, and the values of $\epsh$ are the same order of magnitude for the two cases. We thus expect comparable results for the resonance-trapping phenomenon.

\section{Analysis of the trapping process} \label{sec:trapping}
The numerical studies carried out to analyze the phenomenology of the trapping process have been performed with the map model of Eq.~\eqref{eq:map} as well as the Hamiltonian of Eq.~\eqref{eq:ham} in order to establish conditions under which the adiabatic resonance trapping for the modulated symplectic map can be described by the analytical results for Hamiltonian systems in a neighborhood of the elliptic fixed point. 
\begin{figure}
\begin{center}
	\includegraphics[trim=1truemm 1truemm 3truemm 2truemm, width=0.57\columnwidth,clip=]{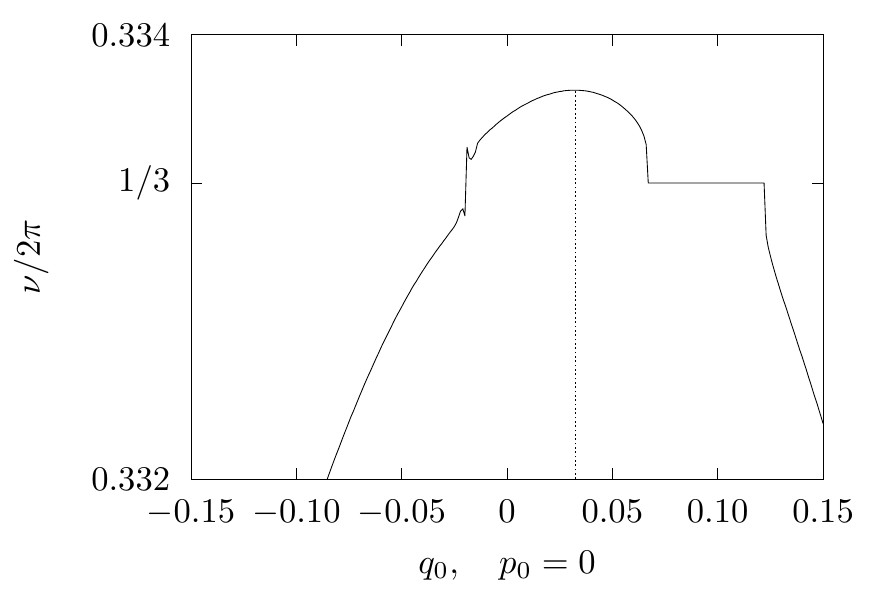}\\
	\includegraphics[trim=1truemm 1truemm 3truemm 2truemm,width=0.57\columnwidth,clip=]{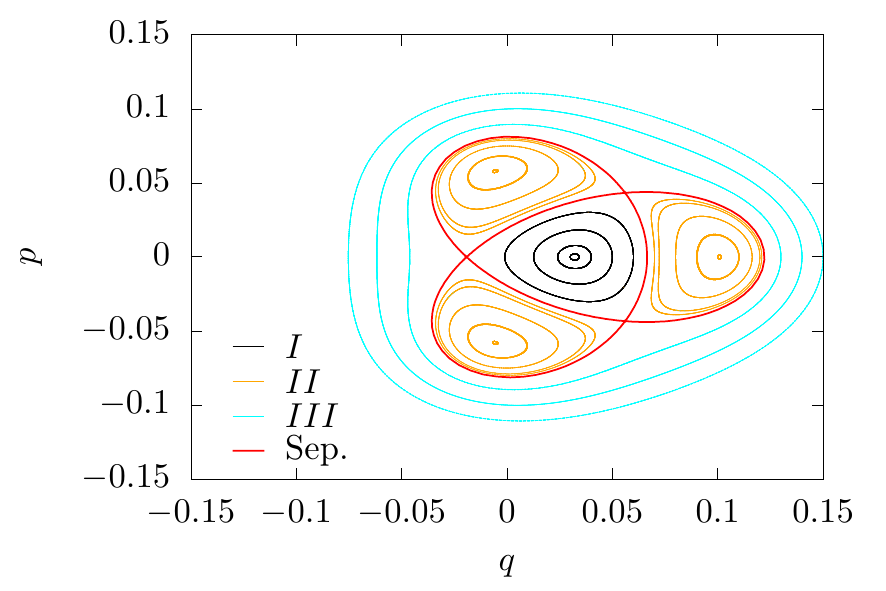}
\end{center}
\caption{Top: Frequency $\omega_0/(2\pi)$ evaluated with a high-accuracy FFT algorithm~\cite{epac,Laskar1999} for an ensemble of orbits with initial conditions $(q_0,0)$ whose orbits are computed using the same model and parameters of the right plot. When the initial conditions are inside the resonance islands, the frequency is locked to the value $1/3$. Bottom: Phase-space portrait of the Poincar\'e map of the Hamiltonian $\ham_{1,3}$~\eqref{eq:ham} evaluated at $\omega \, t=2k\pi,\, k\in\mathbb{N}$, for $\omega_0/(2\pi) = 0.1713$, $k_3=1$, $\epsh=0.28$, $\omega=2.995\,\omega_0$.}
\label{fig:tunephase}
\end{figure}

The main aspect relevant for applications is to investigate which initial conditions are trapped in the resonance and transported in phase space. For this purpose, we determine whether a trajectory is in the resonance islands of the frozen map, relying on the result that the main Fourier component of an orbit in a $n-$resonance island corresponds to the resonant tune $1/n$. We used high-accuracy algorithms for the computation of the main Fourier component~\cite{epac,Laskar1999} to perform the correct identification of the trapped orbits. In Fig.~\ref{fig:tunephase} (top) we show an example of the main frequency for a set of orbits with initial conditions of the form $(q_0,0)$ whose evolution under the Hamiltonian model~\eqref{eq:ham} is evaluated by freezing the time dependence of the system parameters (the corresponding phase-space portrait is shown in the bottom plot of Fig.~\ref{fig:tunephase}). A dependence of the main frequency as a function of  $q_0$ is clearly visible. Note also that the region of constant frequency corresponds to the so-called phase-locking, which occurs when the dynamics is inside a stable island. A sudden jump in frequency can be observed at $q_0=0.066$, which corresponds to an initial condition on the hyperbolic fixed point.

In the following, the concept of trapping fraction will be used in view of studying and qualifying the efficiency of trapping protocols. Given a distribution of initial conditions, the trapping fraction is defined as the ratio of the trapped particles to those in the initial distribution. It is clear that the definition depends on the distribution selected for the initial conditions. For our analysis, it is important to record the original and final regions of the particles: this is made by defining the symbol $\tau_{a \to b}$, where $a$ stands for the region (or regions) from which the initial conditions are taken and $b$ stands for the region in which they are trapped. We remark that the definition of the region from which particles are taken or trapped is based on the phase space topology (such as that visible in Fig.~\ref{fig:tunephase}), at the end of the first stage of the trapping protocol described in the next section. 
\subsection{Hamiltonian models}

To study the phase space of the Hamiltonian of Eq.~\eqref{eq:ham}, it is convenient to use the Poincar\'e map (see the bottom plot of Fig.~\ref{fig:tunephase} for an example of phase-space portrait). When either $\epsh$ or $\omega$ are changed, the separatrices move in phase space changing the enclosed area, while keeping the same topology for $\epsh$ sufficiently small and $\omega$ sufficiently close to the resonance. To describe the phenomenology, the third-order resonance is selected, but the concepts used can be generalized to any resonance order.

According to~\cite{neish1975}, when the system  parameters are adiabatically modulated, the trapping of the orbits into the stable islands and the adiabatic transport are possible. To optimize the trapping probability, we propose a protocol divided into two steps. In the first one, the perturbation frequency $\omega$ is kept constant at a value $\omega_\text{i}< m \, \omega_0$, near the $m$th-order resonance, while the exciter is slowly switched on, increasing its strength $\epsh$ from $0$ to the final value $\eps_\text{h,f}$. In the second stage, the exciter strength is kept fixed at $\eps_\text{h,f}$, and the frequency is modulated from $\omega_\text{i}$ to $\omega_\text{f}$. Both modulations are performed by means of a linear variation in $N$ time steps.

\begin{figure}[htb] 
	\begin{center}
		\includegraphics[trim=1truemm 1truemm 2truemm 2truemm,width=0.57\columnwidth,clip=]{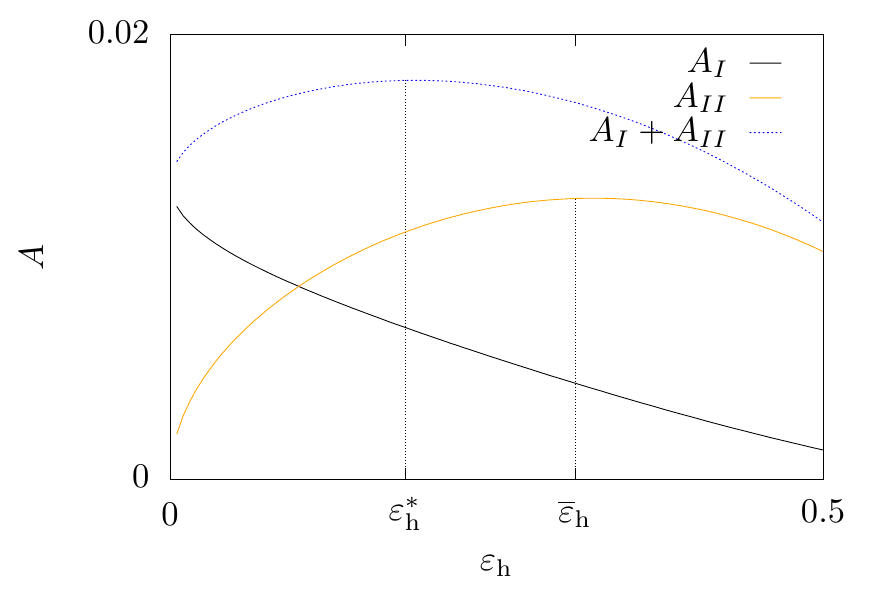}
		\includegraphics[trim=1truemm 1truemm 2truemm 2truemm,width=0.57\columnwidth,clip=]{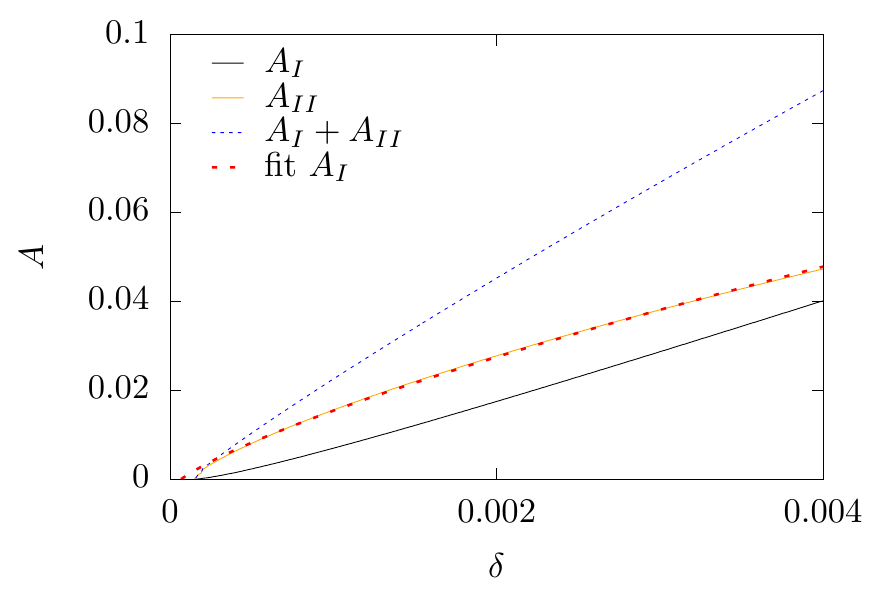}
		\end{center}
        \caption{Top: Areas of Region~I (black), Region~II (orange) and of their sum (blue, dotted) as a function of the exciter strength $\epsh$, for $k_3=1$, $\omega_0/(2\pi)=0.17133$, $\omega=2.995\,\omega_0$ for $\ham_{1,3}$. The sum of the areas has a maximum at $\epsh^*$ whereas $A_\mathrm{II}$ has a maximum at a different point $\overline\eps_\text{h}$. Bottom: Areas of Region~I (black), Region~II (orange) and of their sum (blue, dotted) as a function of the distance from the resonance $\delta=3\,\omega_0-\omega$ with $k_3=1$, $\omega_0/(2\pi)=0.17133$, $\epsh=0.28$ for the Hamiltonian $\ham_{1,3}$. $A_\mathrm{II}$ is fitted as $\alpha\delta^{3/4}+\beta$ (red, dashed line) according to the predictions in~\cite{yellowreport}.}
	\label{fig:areas}
\end{figure}

In the first step, as $\epsh$ increases, the area of the resonance islands increases, thus trapping all orbits that cross the separatrix according to Eq.~\eqref{eq:neish}. The phase space can be divided in three regions (see Fig.~\ref{fig:tunephase}, bottom): the inner region (Region~I) encloses the origin and is limited by the inner part of the separatrices, the resonance region (Region~II) made of the stable islands, and the outer region (Region~III) from the outer part of the separatrix to infinity. The areas of Region~I and~II are shown in Fig.~\ref{fig:areas} as a  function of the exciter strength $\epsh$ (top) and the distance from the resonance $\delta=3\, \omega_0-\omega$ (bottom).

We remark that $A_\mathrm{I}$ is always decreasing with $\epsh$, whereas $A_\mathrm{II}$ is increasing up to $\epsh=\overline\eps_\text{h}$. Therefore, since $A_\mathrm{I}+A_\mathrm{II}$ has a maximum at $\epsh^*<\overline\eps_\text{h}$, the Region~III area is increasing when $\epsh> \epsh^*$.

\begin{figure*}
    \begin{center}
        \includegraphics[trim=2truemm 1truemm 4truemm 2truemm,width=.325\textwidth,clip=]{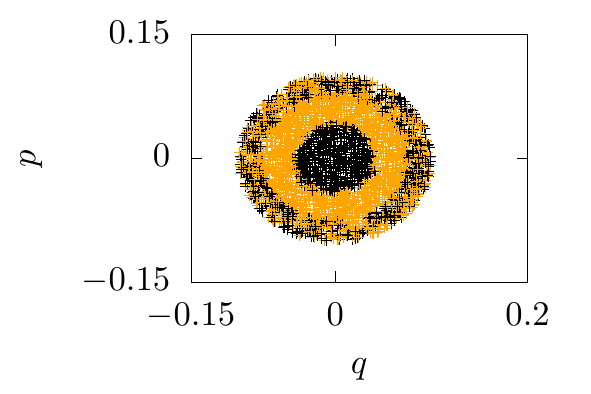}
        \includegraphics[trim=2truemm 1truemm 4truemm 2truemm,width=.325\textwidth,clip=]{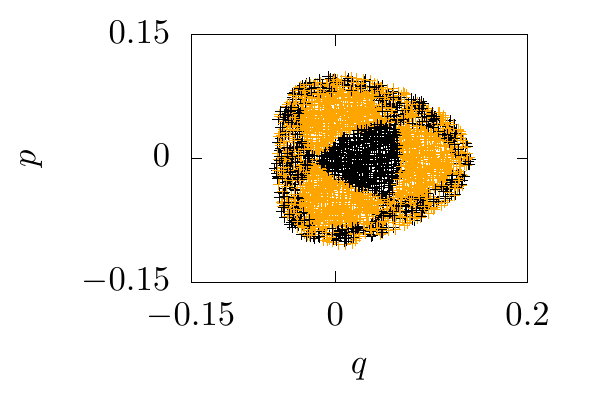}
        \includegraphics[trim=2truemm 1truemm 4truemm 2truemm,width=.325\textwidth,clip=]{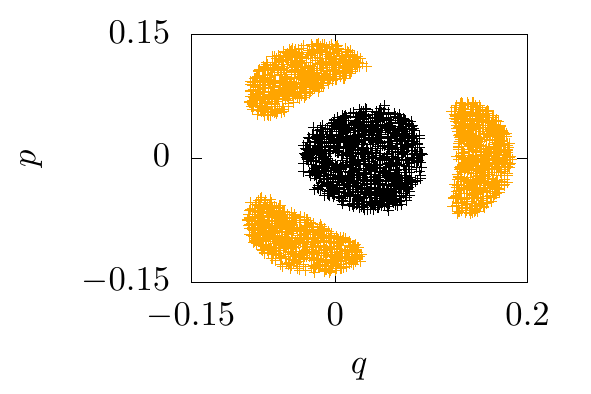}\\
        \includegraphics[trim=2truemm 1truemm 4truemm 2truemm,width=.325\textwidth,clip=]{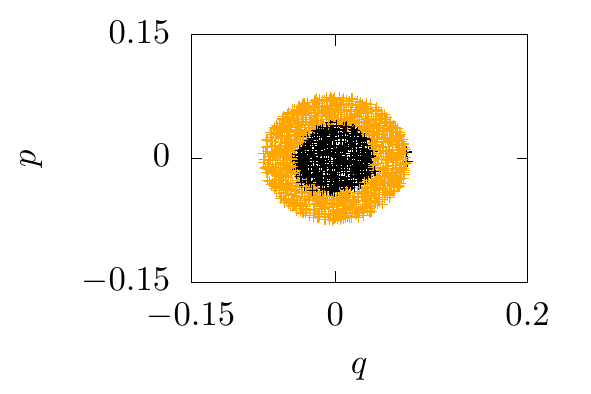}
        \includegraphics[trim=2truemm 1truemm 4truemm 2truemm,width=.325\textwidth,clip=]{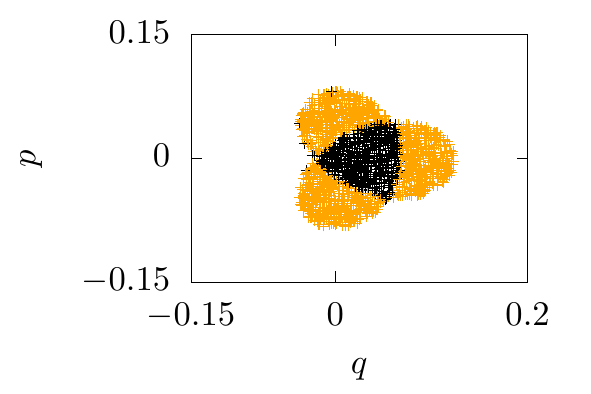}
        \includegraphics[trim=2truemm 1truemm 4truemm 2truemm,width=.325\textwidth,clip=]{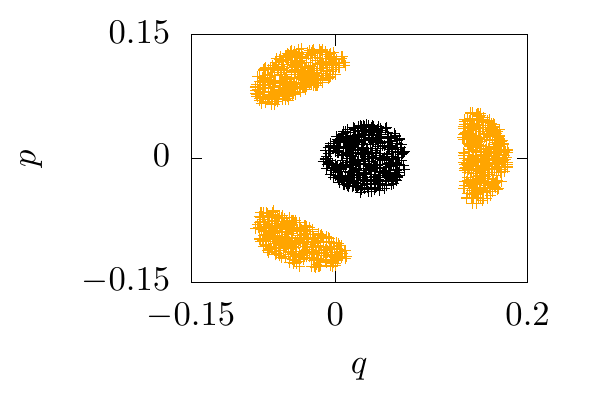}
    \end{center}
\caption{Evolution of an ensemble of particles in  phase space with the colours used to identify in which region each initial condition has been trapped into (Region~I, black, and Region~II, orange) for the Hamiltonian model~\eqref{eq:ham} with $\ell=1$, $m=3$ at the beginning of the process (left column), at the end of the $\epsh$ variation (mid column) and at the end of the frequency variation (right column). The difference between the top and bottom cases is the extent of the distribution of initial conditions: in the top case, some initial conditions are in Region~III, while this is not the case for the bottom case. At large amplitudes, initial conditions can be either trapped in Region~I or~II in the top case, while this is absent in the bottom case. Parameters: $k_3=1$, $\omega_0/(2\pi) = 0.17133$, $\omega_\text{i} = 2.995\,\omega_0$, $\omega_\text{f} = 2.983\,\omega_0$, $\eps_\text{h,f} = 0.28$.}
\label{fig:inmidfin}
\end {figure*}

In the case of an ensemble of initial conditions chosen in Region~I, for $\epsh<\epsh^*$ adiabatic theory ensures that every orbit crossing the inner separatrix is trapped in the resonance, \ie in Region~II. When $\epsh^*<\epsh<\overline\eps_\text{h}$, as both  Region~II and~III areas are increasing, a fraction of orbits will enter into Region~III according to Eq.~\eqref{eq:neish}. These observations are essential for engineering the variation of the system parameters in order to control the trapping and transport phenomena, which is essential for devising successful applications. An example of the behavior described above is shown in Fig.~\ref{fig:inmidfin} in which the evolution of a set of initial conditions under the dynamics generated by $\ham
_{1,3}$ using the protocol for trapping and transport described above is shown. 

Both rows show the evolution of an ensemble of initial conditions under the same dynamics generated by $\ham_{1,3}$ and the colors are used to indicate which region the initial conditions are trapped into. The trapping and transport phenomena are clearly visible, thus indicating that the proposed protocol works efficiently. Between the two rows, the distribution of initial conditions is changed. In the top row, the larger amplitude of the initial conditions is such that an annulus exists in which initial conditions can be trapped either in Region~I or~II. On the other hand, the smaller extent of the initial distribution in the bottom row removes this phenomenon and there is a clear separation between particles that will be trapped in Region~I or~II. We remark also that the initial conditions at large amplitude in the top row contribute to a larger surface of the transported islands and core. 
\begin{figure}[htb]
    \begin{center}
       \includegraphics[width=0.7\columnwidth]{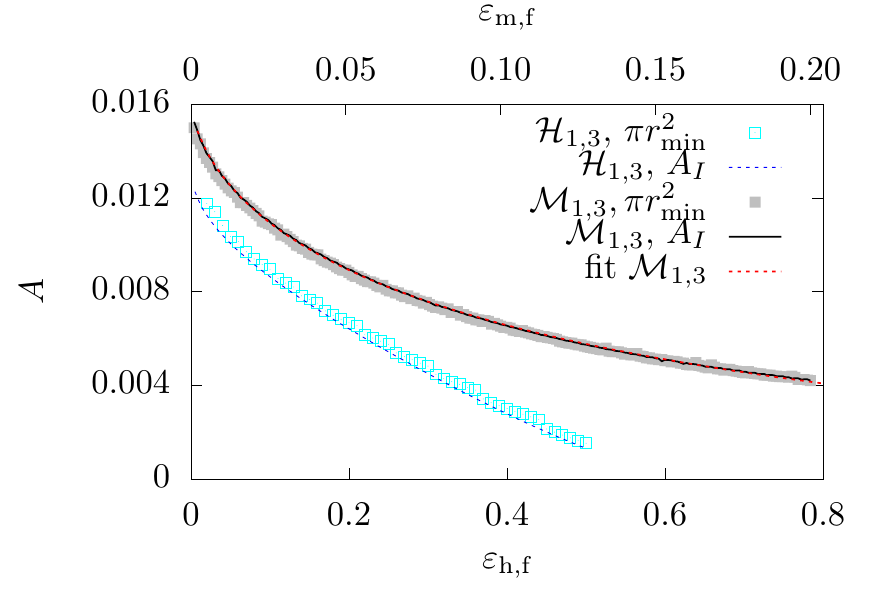}
    \end{center}
\caption{Comparison of the minimum enclosed area of trapped orbits and $A_\mathrm{I}$ at the end of the first stage, \ie the linear variation of the exciter strength, as a function of $\eps_\text{h,f}$, for $\ham_{1,3}$ with $k_3=1$, $\omega_0/(2\pi)=0.17133$, $\omega=2.995\,\omega_0$, and for the corresponding map $\mathcal{M}_{1,3}$. A model o $A_I(\epsm)=a +f(\epsm,b,\eps_0)\epsm^{-2/3}$ fit for $\mathcal{M}_{1,3}$ is presented, where $f(\epsm,b,\eps_0)$ is a form factor ($f(\epsm,b,\eps_0) = b/[1+(\eps_0/\epsm)^{2/3}]$). The dynamics of $\ham_{1,3}$ has been simulated using $N=10^6$ time steps.}
\label{fig:area_rmin}
\end{figure}

According to adiabatic theory~\cite{neish1975}, all particles whose orbit encloses an area $A$ at the end of the first stage satisfying $A_\text{I}(\eps_\text{h,f})<A<A_\text{II}(\min(\epsh^*,\eps_\text{h,f}))$ will be trapped in the resonance, whereas the particles with $A<A_\text{I}(\eps_\text{h,f})$ will remain in Region~I. Furthermore, assuming that the orbits in the excluded region are very close to the origin, we can estimate the average distance from the origin at which the resonance trapping occurs by $r_\text{min}=\sqrt{A_\text{I}(\eps_\text{h,f})/\pi}$. Fig.~\ref{fig:area_rmin} shows the remarkable agreement between the minimum enclosed area of the trapped orbits and the final area $A_\text{I}$ as function of $\eps_\text{h,f}$.
	
If $\eps_\text{h,f}>\epsh^*$, the particles that enclose an area $A$ satisfying $A_\text{II}(\eps_\text{h,f})<A<A_\text{II}(\epsh^*)$ will be found in the external Region~III at the end of the first stage. Therefore, the distribution of initial conditions around the origin can be divided in three parts: the part close to the origin that remains in Region~I, the part trapped in the resonance \ie in Region~II, and the part that is or enters into Region~III.

During the second phase, the exciter frequency is varied to move the resonance in phase space and thus performing the adiabatic transport of the trapped initial conditions. As shown in Fig.~\ref{fig:areas} (bottom), both Region~I and~II increase their area so that no further trapping of orbits close to the origin nor any detrapping from the resonance region are expected. Conversely, the orbits in Region~III will enter either Region~II or I according to the probabilities (see Eq.~\eqref{eq:neish}) that are calculated at the time when the separatrix crossing occurs~\cite{neish1975}. 

We remark that for the systems under consideration, the dependence of $A_i$ on the system parameters is so smooth that approximating the $\lambda$ derivatives of $A_i$ at the time of the actual separatrix crossing, which is needed to compute the trapping probabilities, with a finite difference is an excellent approximation.

\begin{figure}[htb]
	\centering
	\includegraphics[trim=1truemm 1truemm 1truemm 2truemm,width=0.7\columnwidth,clip=]{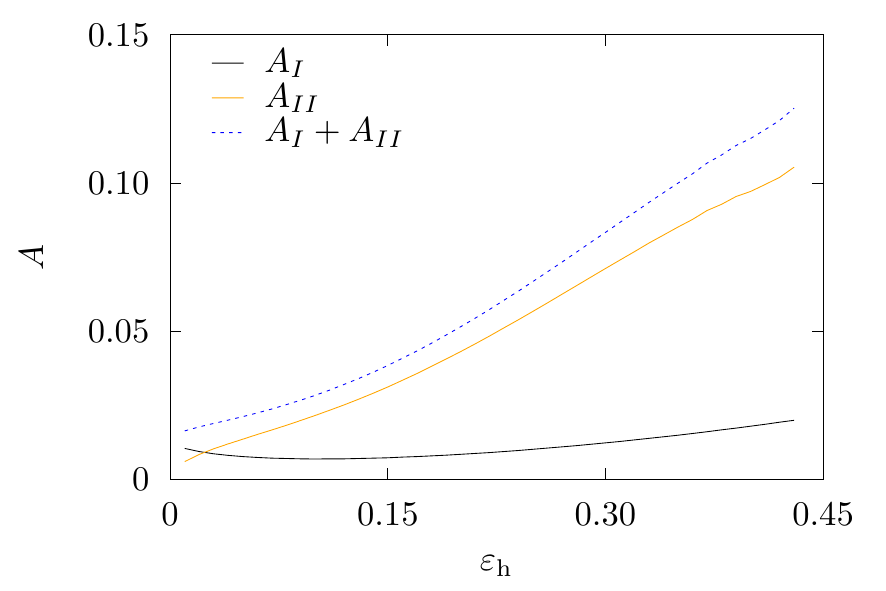}
	\caption{Areas of the center (black), islands (orange) areas, and their sum (blue, dotted) as a function of $\epsh$ with $k_3=1$, $\omega_0/(2\pi)=0.17133$, $\omega=2.995\,\omega_0$ for $\ham_{2,3}$.}
	\label{fig:quad_areas}
\end{figure}

The situation is radically different when one considers $\ham_{2,3}$ as it can be seen in Fig.~\ref{fig:quad_areas}, where the areas of the center (Region~I) and of the islands (Region~II) are shown as a function of $\eps_\text{h}$.

Indeed, in this case, $A_\text{I}$ is decreasing only for a small interval of $\epsh$ around zero, and then it increases, while $A_\text{II}$ increases monotonically, similarly to the sum of the two areas. This implies that there is no possibility for trapping in Region~III. Furthermore, the initial conditions in Region~III will be trapped either in Region~I or~II according to Eq.~\eqref{eq:neish}. 
\subsection{Map models}
\begin{figure}
    \begin{center}
        \includegraphics[trim=1truemm 1truemm 1truemm 2truemm,width=0.57\columnwidth,clip=]{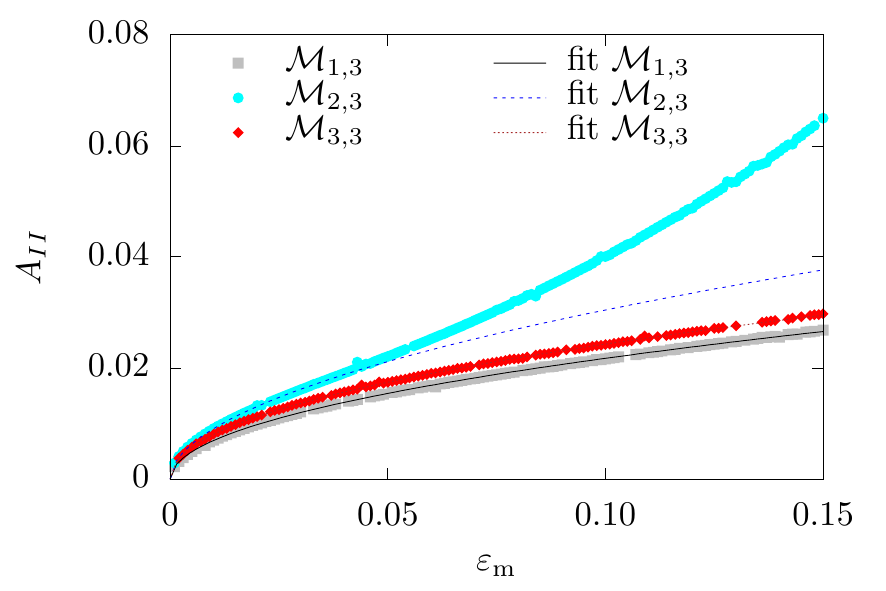}\\
        \includegraphics[trim=1truemm 1truemm 1truemm 2truemm,width=0.57\columnwidth,clip=]{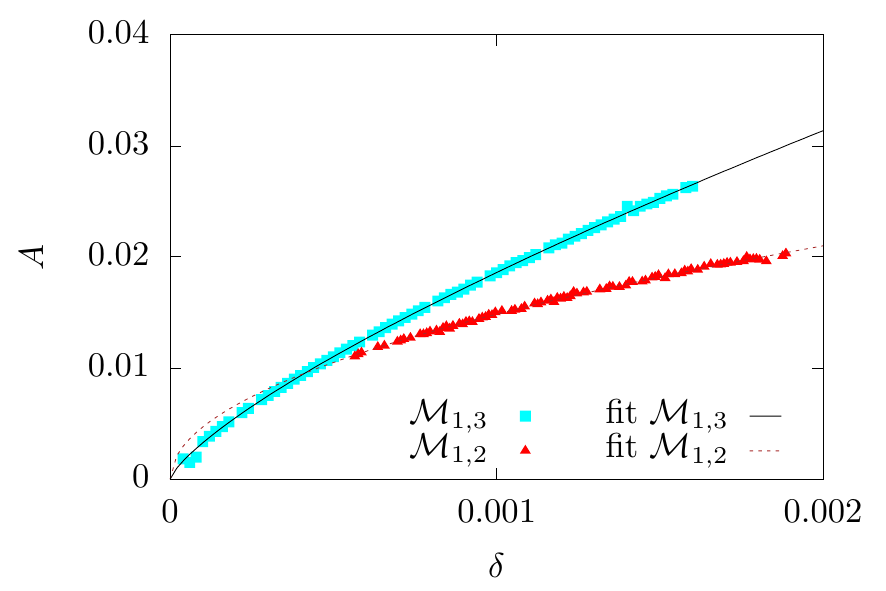}
    \end{center}
\caption{Values of $A_\mathrm{II}$ for $\mathcal{M}_{\ell,m}$ as a function of $\epsm$ (top) and $\delta$ (bottom) and their power-law fits, $A_\text{II}\propto\epsm^{1/2}\delta^{m/4}$, to compare with the prediction in~\cite{yellowreport}. For $\mathcal{M}_{2,3}$, the improved fit from Eq.~\eqref{eq:fitm23} is also shown (top).
\label{fig:areasmap}}
\end{figure}

To investigate the same phenomena using the map~\eqref{eq:map} we need to provide an appropriate framework that allows determining the areas of the various regions as done for the Hamiltonian models.  

We remark that $A_\mathrm{II}$, inspected for different models $\mathcal{M}_{\ell,m}$ follows the scaling laws  $A_\mathrm{II}\propto \eps^{1/2}$ and $A_\mathrm{II}\propto \delta^{n/4}$ outlined in~\cite{yellowreport}. In fact, if we assume that close to a hyperbolic fixed point with action-angle coordinates $(J_\text{h}, \theta_\text{h})$, we can approximate the motion using the pendulum-like Hamiltonian 
\begin{equation}
\ham_{\ell,m} (\overline J, \theta) = \pdv[2]{\ham_{\ell,m}}{J}\eval_{J=J_\text{h}} \frac{\overline J^2}{2} + b J_\text{h}^{m/2}\cos m\, \theta  \, ,
\label{eq:pendulum_ham}
\end{equation}
where $b$ is proportional to $\epsm$, $\overline J = J-J_\text{h}$, and $A_\mathrm{II}$ reads
\begin{equation} 
A_\text{II} = \frac{8}{\left |\pdv[2]{\ham_{\ell,m}}{J}\eval_{J=J_\text{h}} \right |^{1/2}} b^{1/2} J_\text{h}^{m/4} \, ,
\label{eq:Area2}
\end{equation}
and since $J_\text{h}\propto \delta$, we obtain
\begin{equation}
A_\text{II}\propto \epsm^{1/2}\delta^{m/4} \, ,
\label{eq:scaling}
\end{equation}
and this scaling law is shown in Fig.~\ref{fig:areasmap}, where the numerical evaluation of $A_\mathrm{II}$ is compared with the scaling law~\eqref{eq:scaling} as a function of $\epsm$ (top) and $\delta$ (bottom).

Figure~\ref{fig:areasmap} (top) shows clearly that the fit $A_\mathrm{II} = \alpha \epsm^{1/2}$ fails for large values of $\epsm$ for $\mathcal{M}_{2,3}$. We remark that the Hamiltonian of Eq.~\eqref{eq:pendulum_ham} is approximated, hence a better estimate of $A_\mathrm{II}$ can be found by starting from the following Hamiltonian \ie
\begin{equation}
\ham_{\ell,m} (\overline J, \theta) = \pdv[2]{\ham_{\ell,m}}{J}\eval_{J=J_\text{h}} \frac{\overline J^2}{2} + \epsm (J_\text{h}+\overline J)^{3/2}\cos 3\theta \, ,
\end{equation}
and approximating the coefficient of the resonant term with the first-order series expansion in $\overline J$, which gives
\begin{equation}
\ham_{\ell,m} (\overline J, \theta) = \pdv[2]{\ham_{\ell,m}}{J}\eval_{J=J_\text{h}} \frac{\overline J^2}{2} + b J_\text{h}^{3/2} + \frac{3}{2} b J_\text{h}^{1/2} \overline J \cos 3\theta \, .
\end{equation}

The area enclosed by the separatrix is then given by the integral
\begin{equation}
    A_\text{II}=A_{\text{II},0}\int_0^{2\pi}\dd\theta\,\sqrt{k^2\cos^2\theta - \cos\theta + 1} \, , 
    \label{eq:intA2}
\end{equation}
where $A_{\text{II},0}$ is the value of $A_\text{II}$ given in Eq.~\eqref{eq:Area2}, while $k^2=2b/\qty(9\,\Omega_2\, J_h^{1/2})$, \ie $k\propto \epsm^{1/2}$. The expansion of~\eqref{eq:intA2} reads~\cite{stackexchange}
\begin{equation}
A_\text{II}=A_{\text{II},0}\qty(1 + \frac{15\sqrt{2}-5}{24}k^2 - \frac{1}{4}k^2\log k) + O(k^2) \, , 
\end{equation}
which corresponds to a dependence, in $\epsm$
\begin{equation}
A_\text{II}=c_0\epsm^{1/2}(1+ c_1\epsm + c_2\eps_m\log\eps_m) \, , 
\label{eq:fitm23}
\end{equation}
where $c_0$, $c_1$ and $c_2$ can be determined via a fitting process. This estimate, shown in  Fig.~\ref{fig:areasmap} (top) as ``improved fit'', is in good agreement with the $\mathcal{M}_{2,3}$ data.

\begin{figure}[htb]
\centering
    \centering
    \includegraphics[trim=1truemm 1truemm 1truemm 2truemm,width=0.7\columnwidth,clip=]{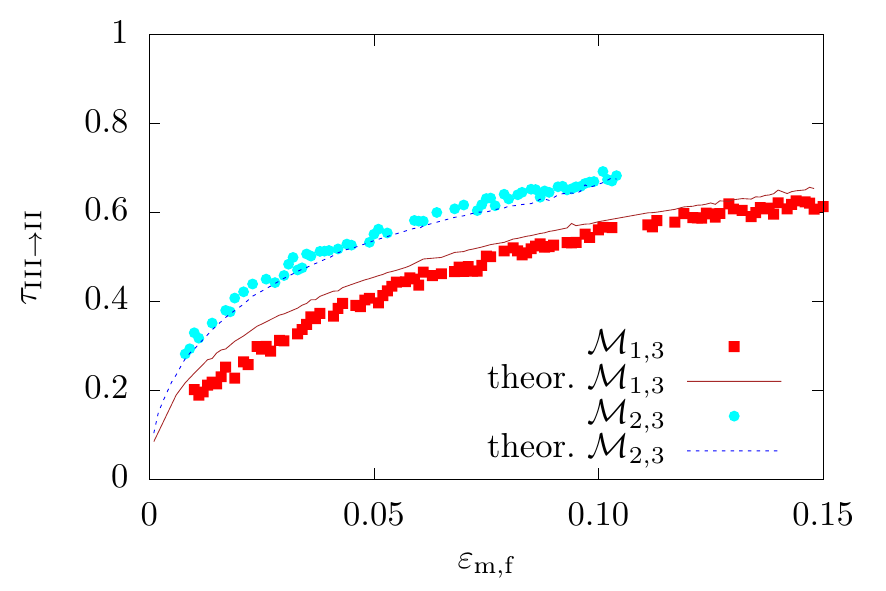}
    \caption{Comparison between the fraction of initial conditions in Region~$\mathrm{III}$ and trapped in Region~$\mathrm{II}$ as a function of $\eps_\text{m,f}$, as computed with numerical simulations (markers) or using Eq.~\eqref{eq:neish} (lines) for $\mathcal{M}_{1,3}$  and $\mathcal{M}_{2,3}$. Note that the derivatives of the areas are estimated considering the finite difference for $\omega_\text{i} = 2.995\,\omega_0$ and $\omega_\text{f}=2.983\,\omega_0$  having set $k_3=1$, $\omega_0/(2\pi) = 0.1713$.}
	\label{fig:neish_map}
\end{figure}

In Fig.~\ref{fig:area_rmin}, we see the excellent agreement between the minimum radius for which the trapping occurs, and the area of Region~I at the end of the first phase of the modulation for the map model, which is a further indication of the validity of the proposed approach. This is also confirmed by the results shown in Fig.~\ref{fig:neish_map}, where the fraction of particles trapped in Region~$\mathrm{II}$ from Region~$\mathrm{III}$ is shown for different $\mathcal{M}_{\ell,m}$ models as a function of $\eps_\text{m,f}$. The predictions from  Eq.~\eqref{eq:neish} are also shown and a very good agreement is observed. Moreover, in Fig.~\ref{fig:area_rmin} we show that a fit function of the form $A_I(\epsm)= a + f(\epsm,b,\eps_0)\epsm^{-2/3}$ fits well the data for $A_\mathrm{I}$, where $f(\epsm,b,\eps_0)$ is a form factor that tends to a constant value for large values of $\epsm$, and that reads
\begin{equation}
    f(\epsm,b,\eps_0) = \frac{b}{1+(\eps_0/\epsm)^{2/3}}
    \label{eq:factorf}
\end{equation}
and this model is fully consistent with the analysis of the minimum trapping action presented in Appendix~\ref{sec:app_rmin}. We remark that the approach presented in the Appendix finds, as estimate of the minimum trapping action, the action of the hyperbolic fixed point. This is proportional to the area of the central region only when it is small \ie at large values of the perturbation parameter $\eps$. For smaller values, the relationship between action of the fixed point and area of the central region is no longer linear, which explains the form of the fit function used.

\section{Comparison of Hamiltonian and map models} \label{sec:simresults}

Extensive numerical simulations have been performed to evaluate the fraction $\tau$ of initial conditions trapped into islands as a function of various parameters both for the map and  Hamiltonian models, for various types of exciters and resonances. The sets of initial conditions are uniformly distributed and are characterized by a maximum radius $R$, i.e. with a p.d.f.\ 
\begin{equation}
    \rho_R(q_0,p_0) = \frac{1}{\pi R^2}  \qquad \text{with } q_0^2+p_0^2 \le R^2 \, .
\end{equation}

\subsection{Case \texorpdfstring{$\ell = 1, m=3$}{l}}

We compare the Hamiltonian dynamics generated by $\ham_{1,3}$, 
whose equations of motion are numerically integrated via the $4$th-order symplectic Candy algorithm~\cite{Candy}, and the map $\mathcal{M}_{1,3}$, for the same scenario where perturbation amplitude and frequency are changed one at a time.
For all our simulations, $k_3=1$, $\omega_0/(2\pi) = 0.1713$. For simplicity, the same number of iterations has been selected to increase linearly the strength of the exciter and its frequency during the modulation stages. The process is implemented with $\omega/(2\pi)=\omega_\text{i}/(2\pi) = 0.5132 = 2.995\,\omega_0/(2\pi)$ and  $\omega_\text{f}/(2\pi) = 0.5112 = 2.983\,\omega_0/(2\pi)$. When not differently stated, we set the number of integration time steps for the Hamiltonian at $N_\text{h}=1.2\times 10^6$, as this value ensures that the modulation is slow enough to achieve adiabatic conditions.

\begin{figure}[htb]
	\begin{center}
	\includegraphics[trim=1truemm 1truemm 1truemm 2truemm,width=0.7\columnwidth,clip=]{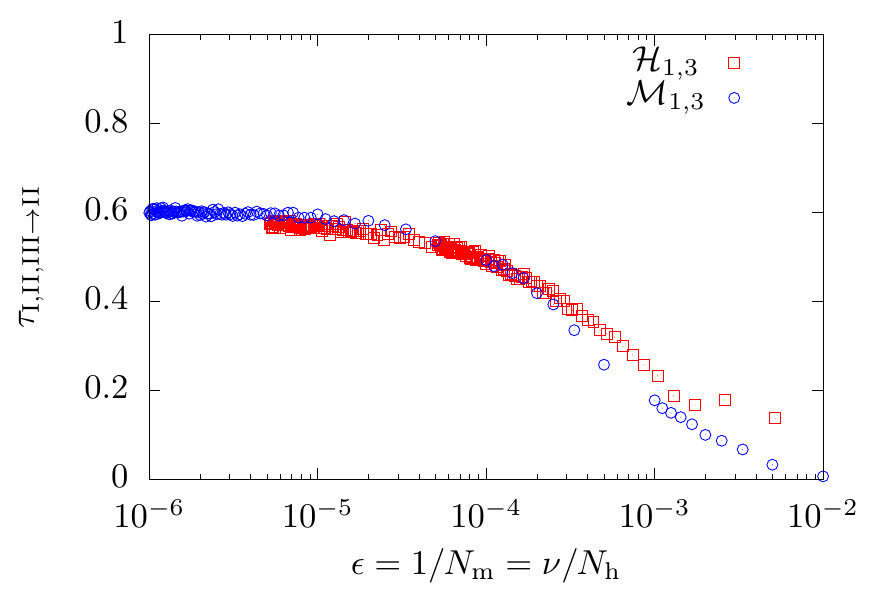}
	\end{center}
	\caption{Trapping fraction as a function of the adiabatic parameter $\epsilon$ for the Hamiltonian $\map_{1,3}$ and $\ham_{2,3}$ for $\omega_0/(2\pi)=0.1713$. As initial condition, a uniform distribution with $R=0.1$ has been used, having set $\epsmf=0.05$ and $\eps_\text{h,f}=\kappa^{3/2}\epsmf = 0.196$. The excellent agreement between map and Hamiltonian models in the adiabatic regime is clearly visible.}
	\label{fig:dip_niter}
\end{figure}

In Fig.~\ref{fig:dip_niter} 
we show the dependence of trapping on the adiabatic parameter $\epsilon=1/N_\text{m}$, where $N_\text{m}$ is the number of iterations of the map, for the $\map_{1,3}$ and $\ham_{1,3}$ models. For the Hamiltonian case, the number $N_\text{h}$ of time steps has been rescaled according to $N_\text{m}=N_\text{h}/\nu$, where $\nu$ is the number of time steps, for $\eps_\text{h}=0$, needed to rotate an initial condition by an angle $\omega_0$ in phase space.

There is a visible excellent agreement in terms of trapping fraction for the map and Hamiltonian models in the adiabatic regime, \ie when $\epsilon \ll 1$, with a slight worsening when $\epsilon$ increases. 

\begin{figure}[htb]
	\centering
		\includegraphics[trim=1truemm 1truemm 1truemm 0truemm,width=0.7\columnwidth,clip=]{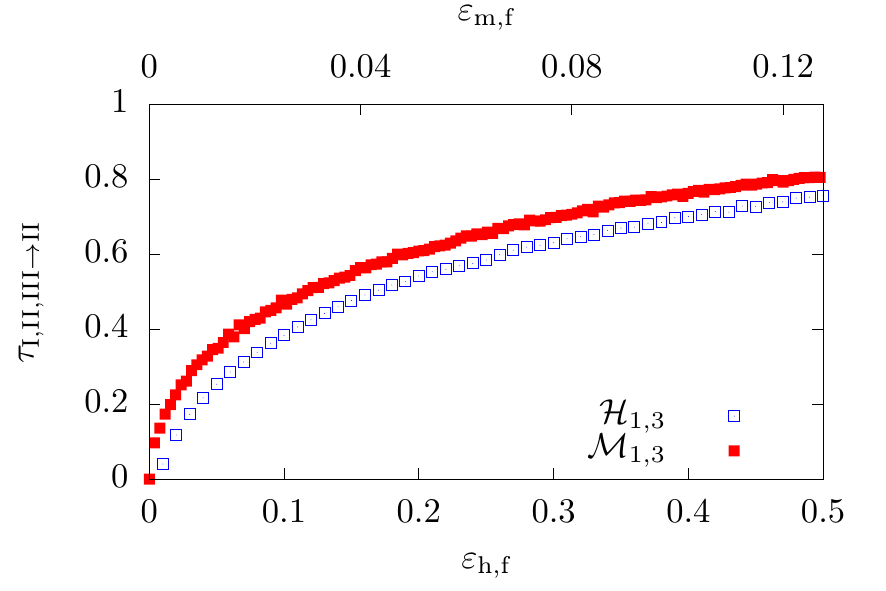}
	\caption{Trapping fraction of the Hamiltonian $\ham_{1,3}$ as a function of $\eps_\text{h,f}$ (red markers) and map $\mathcal{M}_{1,3}$ as a function of $\epsmf$ (blue markers), for initial uniform distribution with  $R=0.1$. The two perturbation strengths are compared via the scaling $\eps_\text{m,f} = \kappa^{3/2} \eps_\text{h,f}$ according to Eq.~\eqref{eq:epsratio}. Note the same functional behavior of the trapping for the two models, which feature only an offset.}
	\label{fig:ham_map_comp}
\end{figure} 

In Fig.~\ref{fig:ham_map_comp} we plot the trapping fraction as a function of $\eps_\text{h,f}$ for the Hamiltonian model and for the map, both for a uniform circular initial distribution with $R=0.1$. The two models can be compared in re-scaling the two perturbation strengths via the ratio $(\omega_2/\Omega_2)^{3/2}$ according to Eq.~\eqref{eq:epsratio}. The graphs describing the evolution of the trapping function are showing the same function dependence on the strength of the exciter, with only an offset between the two curves.

\subsection{Case \texorpdfstring{$\ell = 2, m=3$}{l}} 

Similar studies have been carried out using a quadratic  perturbation $q^2$ in the Hamiltonian, namely

\begin{equation}
\ham_{2,3} = \omega_0 \qty(\frac{p^2}{2}+\frac{q^2}{2}) + \frac{k_3}{3}q^3 + \epsh \frac{q^2}{2} \cos \omega \, t
\end{equation}

and the corresponding map

\begin{equation}
\mathcal{M}_{2,3}:\quad\begin{pmatrix} q_{n+1}\\p_{n+1} \end{pmatrix} = R(\omega_0)\begin{pmatrix}q_n \\ p_n + k_3 q_n^2 + \epsm q \cos \omega \, n \end{pmatrix} \, .
\end{equation}

In this case the mechanism of adiabatic trapping is the same, but the behavior of the areas of phase space regions are quite different (as discussed previously) so that there are important consequences for applications. 

\begin{figure}[htb]
	\begin{center}
	    \includegraphics[trim=1truemm 1truemm 1truemm 0truemm,width=0.7\columnwidth,clip=]{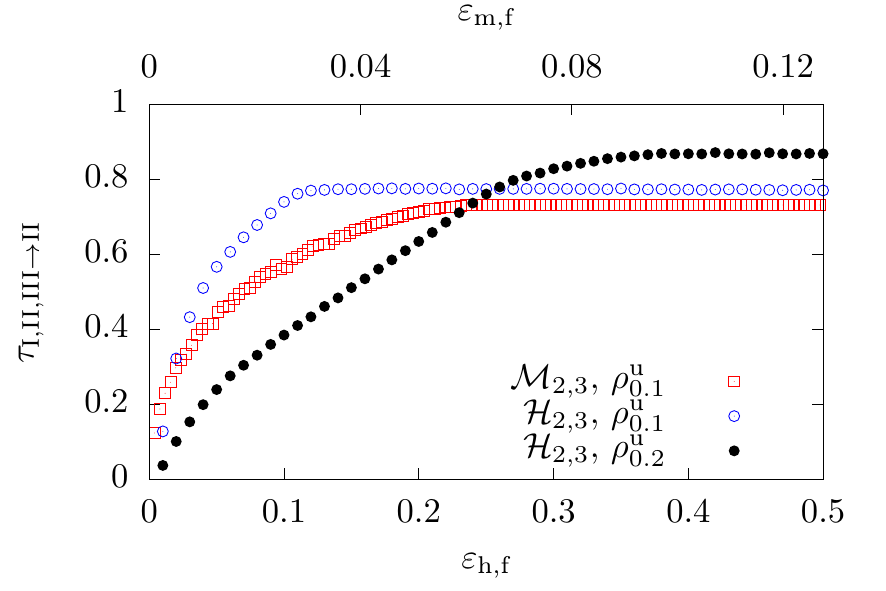}
	\end{center}
\caption{Trapping fraction for $\ham_{2,3}$ as a function of $\eps_\text{h,f}$ (blue and black markers) and $\mathcal{M}_{2,3}$ as a function of $\epsmf$ (red), for initial uniform distributions with $R=0.1$ and $R=0.2$. The two perturbation strengths are compared via the scaling $\eps_\text{m,f} = \kappa^{3/2} \eps_\text{h,f}$ according to Eq.~\eqref{eq:epsratio}.}
\label{fig:ham_map_quad_comp}
\end{figure}

In Fig.~\ref{fig:ham_map_quad_comp} the dependence of the trapping fraction on $\eps_\text{h,f}$ is shown for two values of the radius of the initial uniform distribution. The impact on the trapping fraction is clearly visible. Indeed, if $\eps_\text{h,f}$ is not too small, $A_\text{I}$ and $A_\text{II}$ are both increasing. Hence, if the radius of the initial distribution is reduced, the fraction of particles that remains in Region~I is high. Moreover, after the central area starts growing, trapping into resonance is not possible anymore, and the trapping fraction saturates. The saturation value depends on the radius of the initial distribution and increases for larger values of $R$. Conversely, when $\epsh$ is small, since the resonance islands are created at the origin, an initial distribution with larger radius will place more initial conditions outside of the area swept by the island structure, which prevents them from being trapped. This explains why, for lower values of $\epsh$, we observe a better trapping efficiency for smaller initial distributions. In the same Figure, we show that the map  model presents a qualitatively similar behavior (the scaling of the perturbation strength allows to compare the two models), with a good quantitative agreement observed in the saturation region.

\begin{figure}[htb]
	\begin{center}
	\includegraphics[trim=1truemm 1truemm 1truemm 2truemm,width=0.7\columnwidth,clip=]{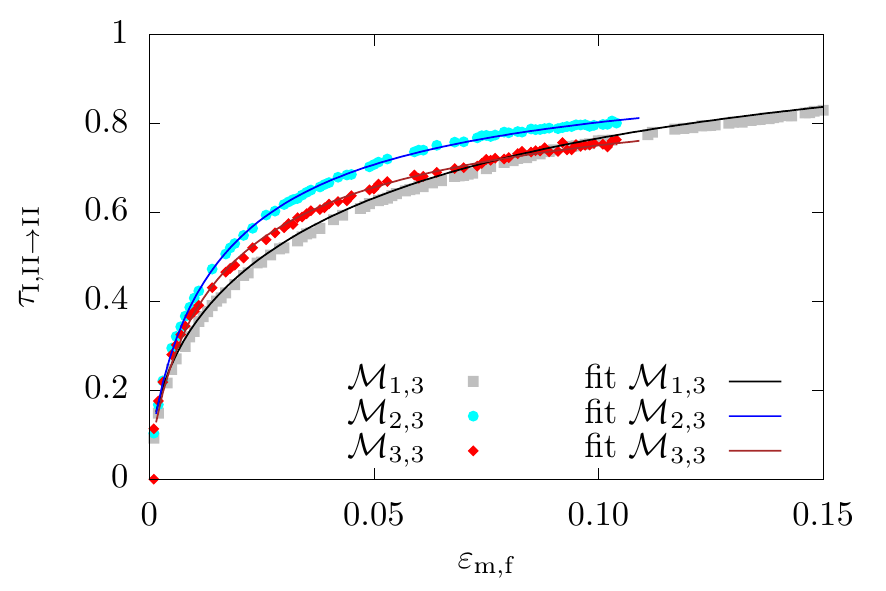}
	\end{center}
	\caption{Trapping fraction for the models  $\mathcal{M}_{\ell,3}$, $\ell=1,2,3$. Fits are presented based on Eq.~\eqref{eq:arearatio}, with $A_\text I(\epsm)=a +f(\epsm,b,\eps_0)\epsm^{-2/3}$ and $A_\text{II}(\epsm) = c\epsm^{1/2}$, in agreement with the models reported in Figs.~\ref{fig:area_rmin} and \ref{fig:areasmap}.}
	\label{fig:compare_poles}
\end{figure}

We have performed numerical studies of the trapping efficiency as a function of $\epsmf$ for a set of initial conditions, selected in Region~I and~II, whose distribution is given by $\rho_R$ with $R = \sqrt{[A_\text{I}(\epsmf)+A_\text{II}(\epsmf)]/\pi}$. In Fig.~\ref{fig:compare_poles} we present a comparison between the trapping fraction computed by means of numerical simulations and the models derived for expressing the surface of Region~I and~II, i.e.
\begin{equation}
    \tau_{\text{I,II}\to \text{II}}  = \frac{A_\text{II}(\epsm)}{A_\text{I}(\epsm) + A_\text{II}(\epsm)}\, ,
    \label{eq:arearatio}
\end{equation}
where $A_\text{I}=a+f(\epsm,b,\eps_0)\epsm^{-2/3}$, the factor $f(\epsm,b,\eps_0)$ having been introduced in Eq.~\eqref{eq:factorf}, and $A_\text{II}=c\, \epsm^{1/2}$. Note that the model presented in Eq.~\eqref{eq:arearatio} is only valid for a uniform initial distribution. For a different radial initial distribution with p.d.f.\ $\rho(r)$ (the angular distribution is assumed to be uniform) we would have
\begin{equation}
    \tau_{\text{I,II}\to \text{II}}  = \frac{\displaystyle\int_{r_\text{I}}^{r_\text{II}}\dd r\, \rho(r)}{\displaystyle\int_{0}^{r_\text{II}}\dd r\, \rho(r)}\, ,
\end{equation}
where $r_i = \sqrt{A_i(\epsm)/\pi}$, $i=\text{I},\text{II}$.

\section{A more complex model} \label{sec:complex}

As a last point, we have considered a more complex model in which an additional parameter has been added, namely a $k_4$ term in the Hamiltonian of Eq.~\eqref{eq:ham} as well as in the map of Eq.~\eqref{eq:map}. 

\begin{figure*}
	\centering
	\includegraphics[trim=2truemm 12truemm 2truemm 5truemm,width=\textwidth,clip=]{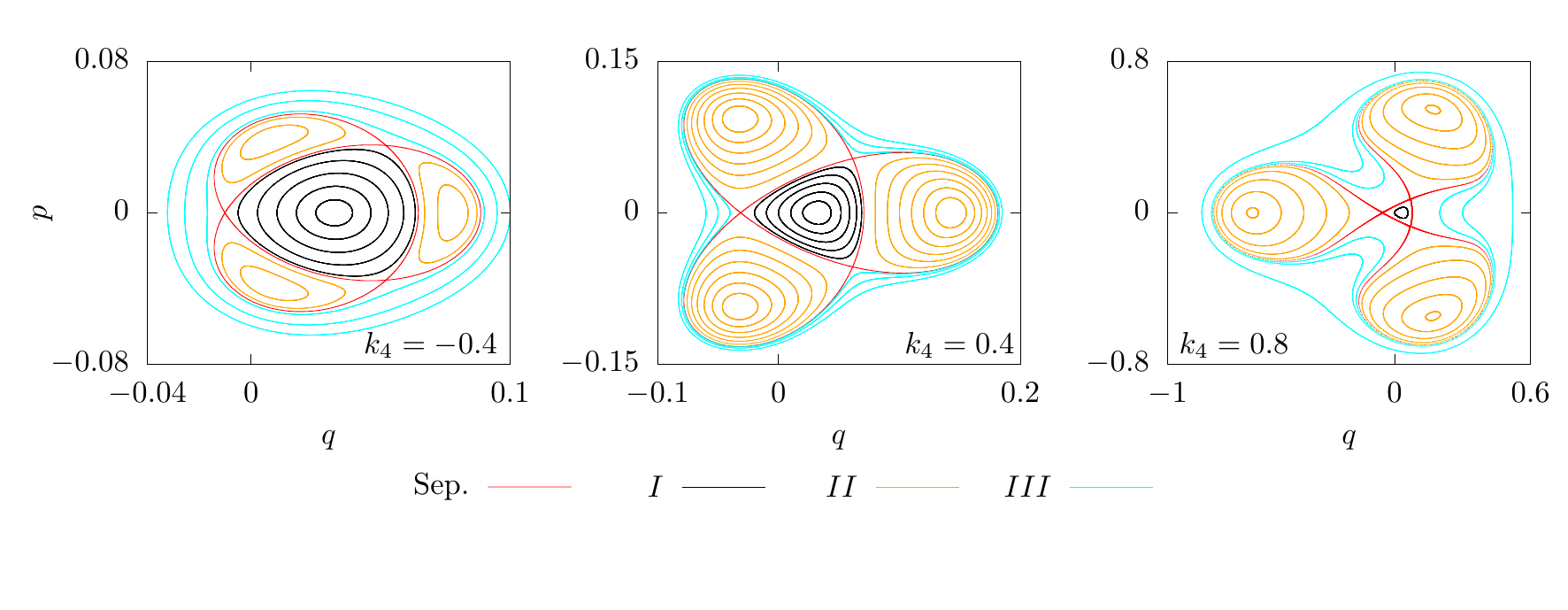}
	\caption{Phase-space portraits for $\ham_{1,3}$ in presence of different quartic nonlinearities $k_4$, having set $\omega_0/(2\pi) = 0.1713$, $\omega=2.995\,\omega_0$, $\eps = 0.28$. The portrait for $k_4=0$ is shown in Fig.~\ref{fig:tunephase} (bottom). Note the difference in scale of the three figures.}
	\label{fig:phsp_k4}
 \end{figure*}

The reason for considering this case is that the phase-space topology changes considerably for different values of $k_4$, as can be seen in Fig.~\ref{fig:phsp_k4}, where three phase-space portraits are shown, corresponding to three values of $k_4$.

Although the global topology is equivalent for all three cases, the detail is not, implying that the surface variation with time of the resonance islands might be rather different between the three cases considered. This would have an important impact on the trapping and transport phenomena. 

\begin{figure}[htb]
	\centering
	\includegraphics[trim=1truemm 1truemm 1truemm 2truemm,width=0.57\columnwidth,clip=]{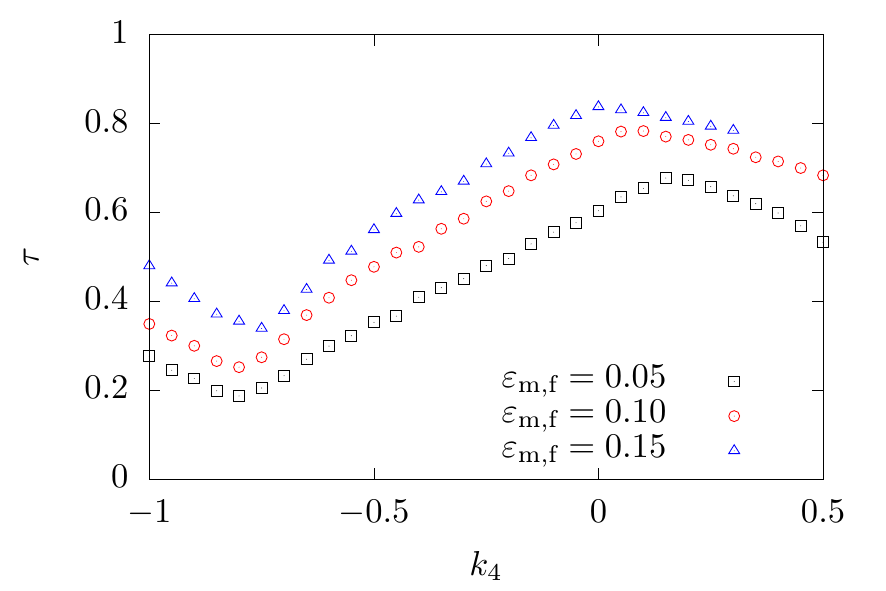}
	\includegraphics[trim=1truemm 1truemm 1truemm 2truemm,width=0.57\columnwidth,clip=]{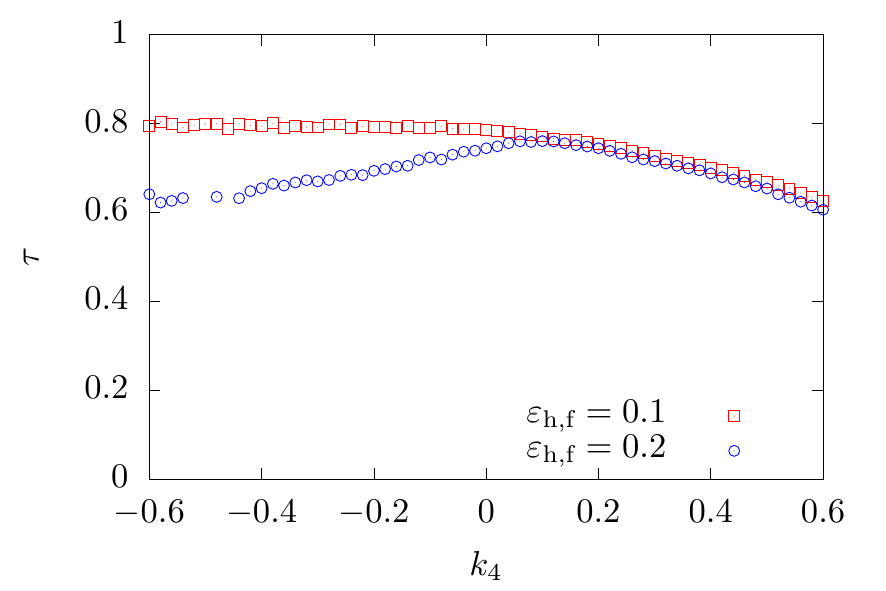}
	\caption{Top: trapping fraction for the model $\map_{1,3}$ as a function of $k_4$. Three cases with different values of $\epsmf$ are shown. Bottom: trapping for the model $\ham_{2,3}$ as a function of $k_4$. Two cases with different values of $\eps_\text{h,f}$ are shown. In all cases, the initial distribution is uniform with $R=0.1$.}
	\label{fig:ham_dip_quad_k4}
\end{figure}

The impact of the $k_4$ term on the trapping fraction has been studied by means of numerical simulations that have been performed on a model of type $\map_{1,3}$ and a second one of type $\ham_{2,3}$ to assess the behavior for different types of time-dependent perturbations. The results are shown in Fig.~\ref{fig:ham_dip_quad_k4}, for the $\map_{1,3}$ (top) and the $\ham_{2,3}$ (bottom) cases.

For the $\map_{1,3}$ model, three cases corresponding to different values of $\epsmf$ have been considered. A strong dependence of the trapping fraction on $k_4$ is clearly visible, with $\epsm$ being also a parameter with a strong impact on trapping.  

For the $\ham_{2,3}$ model, a mild dependence of the trapping fraction on $k_4$ is observed. However, the two cases corresponding to different values of $\eps_\text{2,3}$ behave differently as a function of $k_4$. For $k_4>0$ the two cases feature equal values of the trapping fraction, whereas for $k_4<0$ differences are visible. 

It is worth stressing that even for this more complex model, which shows peculiar features, the trapping and transport processes have been designed using the criteria presented and discussed for the simpler models. This is an indication that the theory, developed for the basic models can also be used to interpret more general cases.

\section{Conclusions} \label{sec:conc}

In this paper, a class of dynamical systems has been considered, in which nonlinear effects are combined with a time-dependent external exciter. This class of systems has been studied both in terms of Hamiltonian as well as using a nonlinear symplectic map. The goals of our analyses were to assess the possibility of deriving effective scaling laws for the efficiency of resonance trapping for adiabatically perturbed symplectic maps using the analytical results of adiabatic theory for Hamiltonian systems, and to identify the application of this class of systems to perform trapping and transport in phase space. Both aspects have been successfully carried out. 

The comparison between Hamiltonian and symplectic map systems has been considered in detail. It has been shown that the adiabatic theory for Hamiltonian systems provides the appropriate framework to describe the trapping and transport phenomena for nonlinear symplectic maps, too. We have shown that adiabatic trapping into stable resonance islands while modulating a periodic, time-dependent perturbation, is an efficient mechanism for phase-space particle transport. A protocol to vary the two system parameters, namely, the strength and frequency of the time-dependent perturbation, has been proposed, which successfully addresses this aspect. The dynamic mechanisms occurring during the separatrix change in phase space have been understood for different models, highlighting the phase-space structure, and comparing the results of the numerical simulations with the theoretical predictions. Several scaling laws have been studied, and extensive simulations have been performed to probe the dependence of trapping and transport features as a function of the systems' parameters. 

The extension of these results to realistic models, as required by physical applications, implies considering multidimensional systems, for which the theory still needs to be fully developed. Nonetheless, the results presented in this paper open a road-map for a feasibility study to apply the resonance trapping induced by an external periodic perturbation in the field of particle accelerators, as a possible improvement of the novel beam manipulations that have been developed to trap beams of charged particles in a circular accelerator.

\section*{Acknowledgments}
We are indebted to Prof.\ A.\ Neishtadt for several discussions and useful suggestions. One of the authors (A.B.) would like to thank the CERN Beams Department for the hospitality during the preparation of this work.

\appendix

\section{Perturbative analysis of an Hamiltonian system with a time-dependent exciter}
\label{sec:appA}

It is convenient to introduce the linear action-angle variables $(I,\phi)$ using $q=\sqrt{2I}\cos\phi$, $p = \sqrt{2I}\sin\phi$, and the Hamiltonian~\eqref{eq:ham} reads
\begin{equation} 
\begin{split}
	\ham_{\ell,m} (I, \phi) = \omega_0 I &+\frac{2^{3/2} k_3}{3} I^{3/2}\sin^3\phi \\ &+ 2^{\ell/2}\epsh I^{\ell/2}\sin^\ell\phi\cos\omega \, t \, .
\end{split}
\label{eq:ham_app}
\end{equation}

We denote by $(J,\theta)$ the action-angle variables of the unperturbed Hamiltonian, \ie for $\epsh=0$, and $J$ turns out to be the adiabatic invariant of the system when no  resonance condition is fulfilled.

To study the adiabatic trapping, we compute the parametric dependence of $(J,\theta)$ from the nonlinear terms using a perturbative approach~\cite{turchetti}.  We apply the Lie transformation $\exp(D_{F(J,\theta)})$ using a generating function

\begin{equation}
F(J, \theta) = k^2_3\sum_{m>2}J^{m/2}f_m(\theta)
\end{equation}
and we obtain a Normal Form Hamiltonian

\begin{equation}
\ham_0(J) = \omega_0 J + \sum_{m>1} \frac{\omega_m(k_3)}{m}J^m \, .
\end{equation}

The perturbative equation for $f_3(\theta)$ reads 
\begin{equation} 
f'_3(\theta) = \frac{2^{3/2}}{3\, \omega_0}\sin^3\theta \, ,
\end{equation}
and it can be integrated, yielding
\begin{equation} 
f_3(\theta) = \frac{2^{3/2}}{3\, \omega_0}\qty(\frac{1}{3}\cos^3\theta - \cos\theta) \, .
\end{equation}

At first order, the change of variables reads

\begin{equation} 
	\begin{aligned}
		I &= J - \pdv{F}{\theta} + \mathcal{O}(J^2) =J-J^{3/2}k^2_3 \, f'_3(\theta) + \mathcal{O}(J^2)\\
		\phi &= \theta + \pdv{F}{J} + \mathcal{O}(J) =\theta+\frac{3}{2}J^{1/2}k^2_3 \, f_3(\theta) + \mathcal{O}(J) \, , 
	\end{aligned}
	\label{eq:transf}
\end{equation}
and the $\omega_2$ coefficient of the Normal Form Hamiltonian is the average value

\begin{equation}
\omega_2 = -\frac{3\omega_0}{2\pi}\int_0^{2\pi}\dd\theta\,f'^2_3(\theta) =-\frac{5\, k^2_3}{6\, \omega_0} \, .
\end{equation} 

The strength of the $m$th-order resonance, $\omega_0+m\, \omega=0$, is given by the $m$th Fourier coefficient of $q^\ell/\ell$, \ie

\begin{equation}
c_{\ell,m} = \frac{1}{2\pi\ell}\int_0^{2\pi}\dd\theta\, e^{im\theta} q^\ell \, ,
\end{equation}
so that to study the third-order resonance ($m=3$) with a linear forcing ($\ell=1$), we can truncate the expansion at $J^{3/2}$, because higher-order terms in $J$ will not have a projection on the Fourier coefficient of $\exp(\pm i3\theta)$. 

The perturbation term is proportional to $q=\sqrt{2I}\sin\phi$ and setting $\eta=2^{3/2}k^2_3/(3\, \omega_0)$ and expanding up to $J^{3/2}$, we obtain

\begin{equation}
\begin{aligned}
q &= \sqrt{2I}\sin\phi \\
&= \sqrt{2} J^{1/2}(1 - J^{1/2}k^2_3 \,  f'_3(\theta))^{1/2}\times \\
& \times \sin\qty(\theta + \frac{3}{2}J^{1/2}k^2_3f_3(\theta)) + \mathcal{O}(J^2) \\
&= \sqrt{2} J^{1/2}(1 -4\eta J^{1/2}\sin^3\theta - \frac{\eta^2}{8} J \sin^6\theta)\times\\
& \times \left [\sin\theta + \frac{J^{1/2}\eta}{2}(\cos^4\theta -3\cos^2\theta)+ \right.\\
& \left.- \frac{\eta^2 J}{8}\sin\theta(\cos^6\theta-6\cos^4\theta+9\cos^2\theta)\right ] + \mathcal{O}(J^2) \, .
\end{aligned}
\end{equation}

The coefficient $[q]_{3/2}$ of the $J^{3/2}$ term reads

\begin{equation}
\begin{split}
[q]_{3/2} & = -\frac{\sqrt{2}\eta^2}{8}\sin\theta \Big( 2\cos^4\theta\sin^2\theta +\\ 
& + \sin^6\theta - 5\cos^6\theta + 9\cos^2\theta\Big) \, , 
\end{split}
\end{equation}
and its Fourier coefficient of order 3 is 
\begin{equation} 
\begin{split}
c_{1,3} & = \frac{1}{2\pi}\int_0^{2\pi}\dd\theta\, e^{3i\theta} [q]_{3/2} =  i\frac{57\sqrt{2}}{256}\eta^2 \\ 
& = i\frac{19\sqrt{2}}{96}\frac{k_3^4}{\omega_0^2} \, .
\end{split}
\end{equation}

Performing analogous computations for the case $\ell=2$ and $m=3$, the term $q^2(\theta,J)$ can be written in the form
\begin{equation} 
\begin{aligned}
        \frac{q^2}{2} & = I \sin^2 \phi \\
        & = J\sin^2\theta + \frac{1}{2}J^{3/2}[q^2]_{3/2}(\theta) +\mathcal{O}(J^2) \, , 
\end{aligned}
\end{equation}
and the Fourier coefficient of $[q^2]_{3/2}(\theta)$ is given by
		
\begin{equation} 
\begin{split}
c_{2,3} & = \frac{1}{2\pi}\int_0^{2\pi}\dd\theta\, e^{3i\theta}\frac{[q^2]_{3/2}}{2} = i\frac{31}{32}\eta \\
& = i\frac{31\sqrt{2}}{48}\frac{k^2_3}{\omega_0} \, .
\end{split}
\end{equation}

By comparing the coefficient $c_{2,3}$ with the corresponding coefficient $c_{1,3}$, one determines the different scale of the perturbation strength in the two cases, namely

\begin{equation} 
\frac{c_{1,3}}{c_{2,3}} = \frac{19}{62} \frac{k^2_3}{\omega_0}
\end{equation}

Finally, we observe that the expansion of $q(J, \theta)$ starts at $J^{1/2}$, whereas the one of $q^2(J, \theta)$ starts at $J$. In general, given a $q^\ell$ perturbation, the lowest-order term is given by $J^{\ell/2}\sin^\ell\theta$. This means that resonances with $m<\ell$ are excited by higher-order perturbation terms, so that the expected relevance for applications is considerably reduced. 

We remark, that the Hamiltonian~\eqref{eq:ham_app} can be analyzed using a different approach. In the new variables one obtains an approximate Hamiltonian of the form

\begin{equation}
    \ham_{1,3}(J,\theta) = \omega_0 J + \frac{\omega_2}{2} J^2 + \epsh c_{1,3}J^{3/2}\cos(3\,\theta - \omega \,t) \, ,
\end{equation}
and we can introduce the slow phase $\gamma = 3\,\theta-\omega \, t$ via a time-dependent generating function $G(\tilde{J},\theta) = \tilde{J}(3\,\theta-\omega \, t)$. Setting $\delta = 3\, \omega_0 -\omega$ as the distance from the resonance, we have

\begin{equation} 
\ham_{1,3}(J,\theta) = \delta \tilde{J} + \frac{9\omega_2}{2} \tilde{J}^2 + \epsh c_{1,3} \, 3^{3/2} \tilde{J}^{3/2}\cos\gamma 
\end{equation}
and defining the parameters 
\begin{equation}
    \lambda = -\frac{4\delta}{9 \, \omega_2} \qquad \qquad \mu = \sqrt\frac{2}{3} \frac{\epsh \, c_{1,3}}{\omega_2}
\end{equation}
one obtains by rescaling the Hamiltonian
\begin{equation}
\ham_{1,3}(\tilde J,\gamma) = (2\tilde J)^2 - \lambda (2 \tilde J) + \mu (2 \tilde J)^{3/2}\cos\gamma \, .
\label{eq:ham_neish}
\end{equation}

The dynamics generated by this Hamiltonian can be studied, for what concerns trapping via separatrix crossing, with the methods exposed in~\cite{neish1975,Neishtadt2013}. Its phase space, in fact, features, depending on $\lambda$ and $\mu$, an hyperbolic point at the crossing of separatrices, which enclose an inner and an outer region.

\section{Analysis of the minimum trapping action}
\label{sec:app_rmin}

From the observations reported in the main body of the article, for any value of $\epsm$ or $\epsh$ the phase-space islands appear at some amplitude, which determines the smallest radius for which particles are trapped into the islands. A simplified approach to determine an estimate for the minimum action starts from the Hamiltonian
\begin{equation}
H(I,\theta,\lambda)=H_0(I,\lambda)+\eps \,  I^{m/2}\cos(m\, \theta-\omega \, t)
\label{eq:app_ham0s}
\end{equation}
that corresponds to a forced  nonlinear oscillator with a resonance condition
\begin{equation}
m\frac{\partial H_0}{\partial I}(I_\mathrm{r},\lambda)-\omega_\mathrm{r}=0 
\end{equation}
that defines the resonant action $I_\mathrm{r}(\lambda)$ (when it is real). Note that it is always possible to introduce the angle $m\, \theta=\phi$ and the re-scaling of the action $J=I/k$, so that the Hamiltonian reads
\begin{equation}
H(J,\phi,\lambda)=H_0(k J,\lambda)+\eps \, k^{m/2} J^{m/2}\cos(\phi-\omega \, t) \, .
\end{equation}

The resonant phase $\gamma=\phi-\omega_\mathrm{r} t$ can be introduced by using the generating function
\begin{equation}
G(J,\phi)=J(\phi-\omega_\mathrm{r} \, t)
\end{equation}
and one obtains the pendulum-like system
\begin{equation}
H(J,\phi,\lambda)=H_0(k J,\lambda)-\omega_\mathrm{r} \, J+\eps \,  k^{m/2} J^{m/2}\cos \gamma \, .
\label{eq:app_pendulum}
\end{equation}

To study the nonlinear resonance crossing, we assume
\begin{equation}
\frac{\partial H_0}{\partial J}-\omega_\mathrm{r}=\Delta \omega_0(\lambda)+\Omega \, J
\end{equation}
so that the resonance amplitude in phase space is given by
\begin{equation}
J_\mathrm{r}(\lambda)=-\frac{\Delta \omega_0(\lambda)}{\Omega}\ge 0 \, .
\end{equation}

We can further reduce the Hamiltonian to that of a forced pendulum by using the generating function
\begin{equation}
F(\hat J,\gamma, \lambda)=\gamma(\hat J+J_\mathrm{r}(\lambda)) \, ,
\end{equation}
and the new Hamiltonian has the form
\begin{equation}
H(\hat J,\gamma,\lambda)=\frac{\Omega}{2} \hat J^2+\eps \, k^{m/2}  (\hat J+J_\mathrm{r}(\lambda))^{m/2}\cos \gamma +\gamma \, \epsilon
J_\mathrm{r}'(\lambda)
\end{equation}
where $\lambda=\epsilon \, t$ ($\epsilon\ll 1$)  and $J_\mathrm{r}'=\dv*{J_\mathrm{r}}{\lambda}$. The condition for the existence of fixed points is
\begin{equation}
    \begin{dcases}
    \frac{\partial H}{\partial \hat J}&=\Omega \hat J+\eps \, k^{m/2} \frac{m}{2} (\hat J+J_\mathrm{r}(\lambda))^{m/2 - 1}\cos \gamma=0\\
    \frac{\partial H}{\partial \gamma}&=-\eps \, k^{m/2} (\hat J+J_\mathrm{r}(\lambda))^{m/2}\sin \gamma+\epsilon J_\mathrm{r}'(\lambda)=0 \, .
    \end{dcases}
\end{equation}

The first equation provides the resonance position in  phase space, whereas the second one provides a condition on the existence of the resonance since we obtain
\begin{equation}
\qty|\sin \gamma|=\frac{\epsilon}{\eps} \frac{|J_\mathrm{r}'(\lambda)|}{k^{m/2} J_\mathrm{r}(\lambda)^{m/2}}
\end{equation}
and $\qty|\sin\gamma|\le 1$. We observe that for $\epsilon\ll 1$ (adiabatic parameter) we have the existence of the resonance for small values of the actions $J_\mathrm{r}(\lambda)$. However, for fixed $\epsilon/\eps$ ratio we obtain a condition for the resonance as
\begin{equation}
J_\mathrm{r}(\lambda)^{m/2}\ge C \frac{\epsilon}{\eps} \, ,
\end{equation}
where $C$ is a suitable constant, which means the existence of a minimal trapping action $J_\text{min}$ that scales as
\begin{equation}
J_\text{min} \propto \left (  \frac{\epsilon}{\eps} \right )^{2/m}
\label{eq:app_jmin}
\end{equation}
and, \eg for $m=3$ then $J_\text{min} \propto \eps^{-2/3}$.

\bibliography{biblio}
\bibliographystyle{unsrt}

\end{document}